\documentclass[twocolumn,runningheads]{svjour2}
\smartqed  % flush right qed marks, e.g. at end of proof
\usepackage{graphicx}
\usepackage{mathptmx}   % use Times fonts if available on your TeX system
\usepackage[authoryear]{natbib}
\journalname{Astrophysics and Space Science (CoRoT/ESTA Volume)}
\begin{document}

\title{CL\'ES, Code Li\'egeois d'\'Evolution Stellaire}
\author{R.~Scuflaire \and S.~Th\'eado \and J.~Montalb\'an \and A.~Miglio
  \and P.-O.~Bourge \and M.~Godart \and A.~Thoul \and A.~Noels}
% \institute{R.~Scuflaire \at
%   Institut d'Astrophysique et de G\'eophysique, Universit\'e de Li\`ege, all\'ee
%   du 6 Ao\^ut 17, B-4000 Li\`ege, Belgium \\
%   Tel.: +32-4-3669730\\
%   Fax: +32-4-3669737\\
%   \email{R.Scuflaire@ulg.ac.be}
\institute{R.~Scuflaire \and S.~Th\'eado \and J.~Montalb\'an \and A.~Miglio
  \and P.-O.~Bourge \and M.~Godart \and A.~Thoul \and A.~Noels
  \at
  Institut d'Astrophysique et de G\'eophysique, Universit\'e de Li\`ege, all\'ee
  du 6 Ao\^ut 17, B-4000 Li\`ege, Belgium
% \email{R.Scuflaire@ulg.ac.be}
}

\date{Received: date / Accepted: date}

\maketitle
%*******************************************************************
\begin{abstract}
Cl\'es is an evolution code recently developed to produce stellar models meeting
the specific requirements of studies in asteroseismology. It offers the users a
lot of choices in the input physics they want in their models and its
versatility allows them to tailor the code to their needs and implement easily
new features. We describe the features implemented in the current version of the
code and the techniques used to solve the equations of stellar structure and
evolution. A brief account is given of the use of the program and of a solar
calibration realized with it.
\end{abstract}

%-------------------------------------------------------------------
\section{Introduction}\label{intro}

Since the early 70s, the asteroseismology group at the Institute of Astrophysics
and Geophysics of Li\`ege has been using an evolution code derived from the
Henyey code. This code has been continuously updated. Nevertheless, with the
progress of asteroseismology, it became clear that the frequencies and the
stability of the oscillation modes were extremely sensitive to details of the
model, which were unimportant for the computation of stellar evolution. A few
years ago, it was thus decided to write a new code, meeting the specific
requirements of our studies in asteroseismology. This code has been named
\textit{Cl\'es}, the acronym of \textit{Code Li\'egeois d'\'Evolution
Stellaire}.  It is still in an active phase of development. A number of persons
have contributed and contribute to its development. The main objective of this
effort is to have a versatile stellar evolution code producing models precise
enough to be useful for studies in asteroseismology. We try to maintain a clear
structure so that every user will be able to tailor the code to his own needs. 

Cl\'es has been developed on a GNU/Linux system (more precisely a Slackware
one). No effort has been made to port the software to another platform. It is
written in FORTRAN 77. A few relatively standard extensions of the language
provided by the GNU Fortran compiler g77 have been used. A few scripts use the
Bash shell and the script building the Makefile is a Tcl script.

The present version of the code is numbered 18. The code is not unique. As it is
easy to customize, several versions of the code have been developed from the
standard version, implementing different features needed by the works already
realized or in progress. The most interesting of these will be included in the
next standard version of the code so as to preserve its uniqueness. The features
described below do not necessary belong to the standard version. After a brief
reminder of the equations we have to solve, we describe the physical
assumptions. Two sections are then devoted to technical aspects and we end with
considerations on the use of Cl\'es, a solar calibration and a brief discussion.

Due to limitations in the physics already implemented, the present version of
Cl\'es is not intended for the helium burning phases of the evolution.

%-------------------------------------------------------------------
\section{Structure and evolution equations}\label{equations}

\subsection{Structure}

The quasi-equilibrium structure of a star is governed by the following
equations (we use well-known notations):
\begin{eqnarray}
\frac{\partial P}{\partial r}&=&-\frac{Gm\rho}{r^2}\,,\\
\frac{\partial m}{\partial r}&=&4\pi r^2\rho\,,\\
\frac{\partial L}{\partial r}&=&4\pi r^2\rho\left(\epsilon+\epsilon_g\right)\,,
  \label{eq:dL/dr}\\
\frac{\partial T}{\partial r}&=&-\frac{Gm\rho T}{r^2P}\nabla\label{eq:dT/dr}\,.
\end{eqnarray}

In (\ref{eq:dL/dr}), the luminosity $L$ does not include the neutrino
luminosity and the term $\epsilon$ denotes the rate of production of nuclear
energy from which we have already excluded the neutrinos losses.
In the same equation, the
term $\epsilon_g$ deserves a comment. It hides a time derivative and governs the
evolution rate when no significant nuclear energy sources are available, for
instance in the beginning of the pre-main sequence phase. It reads
\begin{equation}
\epsilon_g=-\frac{dU}{dt}+\frac{P}{\rho^2}\frac{d\rho}{dt}
  =-T\frac{dS}{dt}-\sum_i\mu_i\frac{dn_i}{dt}\,.
\end{equation}
where $\mu_i$ is the chemical potential of element $i$ and $n_i$ the number of
moles per gram. In this expression, some people keep only the term $-TdS/dt$
\citep{K1990}, but this neglects the effects of changes in chemical composition.
An enlightening discussion of the correct form of this term is given by
\citet{S1970}.

In (\ref{eq:dT/dr}), $\nabla$ stands for $d\ln P/d\ln T$ and is
computed differently according to whether the considered point belongs to a
radiative or convective zone. In a radiative zone, the diffusion approximation
allows us to write the temperature gradient as
\begin{equation}
\nabla=\nabla_{rad}=\frac{3\kappa LP}{16\pi acGmT^4}\,.
\end{equation}
The expression of $\nabla$ in a convective zone is discussed in the section
devoted to convection.

\paragraph{Boundary conditions.}

The differential equations of structure must be supplemented with boundary
conditions. The conditions at the center are obvious, $m$ and $L$ must vanish as
$r^3$. In the external layers of the star, the diffusion approximation
progressively breaks down as the optical depth approaches unity. Rather than
integrating the full radiative problem, it is usual to join smoothly the
interior solution to a precomputed atmosphere model, which imposes, in this way,
external boundary conditions on the interior model.

\subsection{Chemical evolution}

The equations governing the rate of change of the abundances of the elements may
be written in the form
\begin{equation}
\frac{dX_i}{dt}=\sum\limits_j R_{ij}
-\frac{1}{\rho r^2}\frac{\partial}{\partial r}
\left(r^2\rho X_iw_i\right)\,.
\label{eq:ChemEvol}
\end{equation}
$R_{ij}$ expresses the contribution of the $j$-th nuclear reaction to the
variation of the species $i$. It depends on the abundances of several species in
a non linear way and needs no further explanation.  $w_i$ is the diffusion
velocity of element $i$. The computation of this term is described in section
\ref{diffusion}.

In convective and overshooting zones, the different\break species are rapidly
mixed.  In a few evolution codes this rapid mixing is described as a diffusive
process with appropriate coefficients. In Cl\'es, convective mixing is an
instantaneous process and the chemical homogeneity of convective zones is
imposed. In a given convective zone, the abundances $X_i$ do not depend on $r$
and (\ref{eq:ChemEvol}) is replaced by its integrated form over the
entire convective zone:
\begin{eqnarray}
m_c\frac{dX_i}{dt}=&&\int_{m_1}^{m_2}\sum_j R_{ij}\,dm\nonumber\\
&&-(4\pi r^2\rho X_iw_i)_2+(4\pi r^2\rho X_iw_i)_1\,,
\label{eq:ChemEvolConv}
\end{eqnarray}
where $m_c$ is the mass of the convective zone and indexes 1 and 2 refer to the
bottom and the top of the convective zone respectively.

\paragraph{Initial conditions.}

The first model of a sequence is always on the Hayashi track. It is chemically
homogeneous and its central temperature is between $4\times10^5$ and
$5\times10^5$~K.
The initial chemical composition is defined by $X$ and $Z$, the mass fractions
of hydrogen and metals. The metal mixture and the isotopic abundances can be
changed by the user. It is however his responsibility to select appropriate
opacity tables. The standard version of Cl\'es comes with two metal mixtures,
GN93 \citep{G1993} and AGS05 \citep{A2005} and the \citet{A1989} isotopic
abundances. It must be stressed that for Li, Be and B, contrary to other
elements, we do not use solar photospheric abundances. As these elements may
have undergone nuclear processing at the basis of the solar convective zone, we
have adopted their meteoritic abundances.

%-------------------------------------------------------------------
\section{Input physics}\label{physics}

\subsection{Equation of state}
\label{sec:eos}
We do not compute the equation of state (EOS) in the evolution code. A few EOS
are provided in the form of pre-computed tables which are interpolated.
We are particularly careful with respect to the consistency between the
different thermodynamic quantities provided by the equation of state. The
independent variables are $\log\rho$, $\log T$, $X$ and $Z$ ($\log$ stands for
the decimal logarithm). At each grid point, we store only $\log P$, $\log C_v$,
$P_\rho=(\partial\log P/\partial\log\rho)_T$ and $P_T=(\partial\log
P/\partial\log T)_\rho$ for the gas.  In the EOS routine, these quantities are
interpolated, the easily computed contribution of the radiation is taken into
account and all other thermodynamic quantities are deduced from well-known
identities. The contributions of the radiation are included as follows.
\begin{eqnarray}
&&P=P_{gas}+P_{rad}\quad\mbox{with}\quad P_{rad}=\frac{1}{3}aT^4\,,\\
&&C_v=C_{v,gas}+\frac{4aT^3}{\rho}\,,\\
&&P_\rho=\beta P_{\rho,gas}\quad\mbox{with}\quad\beta=P_{gas}/P\,,\\
&&P_T=\beta P_{T,gas}+4(1-\beta)\,.
\end{eqnarray}
Adiabatic exponents and $C_p$ are then computed,
\begin{equation}
\Gamma_3-1=\frac{P_TP}{C_v\rho T}\,,
\end{equation}
\begin{equation}
\Gamma_1=P_\rho+(\Gamma_3-1)P_T\,,
\end{equation}
\begin{equation}
C_p=\Gamma_1C_v/P_\rho\,.
\end{equation}

Three EOS tables have been used with Cl\'es, implementing
CEFF \citep{C1992}, OPAL 2001 and OPAL 2005 \citep{R2002}. We have computed the CEFF
table with a routine kindly provided by J.~Christensen-Dalsgaard. A metal
mixture defined by the abundances of ten metals can be input to the routine. As
the EOS is not very sensitive to the metal mixture, we have computed only one
table with a mixture as close as possible to the GN93 mixture. The OPAL 2001
and 2005 EOS
are available in tabular form for a fixed metal mixture of four elements (C, N,
O and Ne). As the $C_v$ from OPAL 2001 tables are known to be inaccurate
\citep{B2003}, $C_v$ has been computed from the other tabulated quantities.

For the necessity of diffusion a routine computes the stage of ionization of all
the elements involved in the diffusion process (Saha's equation).

\subsection{Opacity}

The standard version of Cl\'es uses OPAL opacities \citep{I1996}, completed with
the opacities of \citet{A1994} at low temperature. Both tables have been
smoothly merged into a single table. In the temperature domain $\log T\in
[3.9,4.15]$ where the opacity is defined in both tables, we use an opacity
$\kappa$ defined as the average
\begin{equation}
\log\kappa=(1-\theta)\log\kappa_{AF}+\theta\log\kappa_{OPAL}\,,
\end{equation}
In this expression,
$\theta$ is the unique third degree polynomial in $\log T$, with vanishing
derivatives at both ends of the interval and taking zero and unity values at the
lower and higher ends respectively.

A few opacity tables have been built for different metal mixtures.
Appropriate tools
allow the user to build new tables in the format required by Cl\'es from tables
generated at the OPAL site and the tables of Alexander and Ferguson. For
particular studies, tables using the new \citet{F2005} opacities and OP
opacities \citep{B2005} have also been built \citep{Mi2007}.

For the computation of models with local variations of the metal mixture,
several tables with different
metal mixtures are loaded and interpolated \citep{MiBo2007}.

Up to now the effects of conduction on opacities have been neglected.

\subsection{Nuclear reactions}

The following reactions are included in our nuclear reaction network.

p-p chains:
\begin{eqnarray}
&&2\ {}^1\mbox{H}\rightarrow{}^2\mbox{H}+e^++\nu\\
&&{}^2\mbox{H}+{}^1\mbox{H}\rightarrow{}^3\mbox{He}+\gamma\\
&&2\ {}^3\mbox{He}\rightarrow{}^4\mbox{He}+2\ {}^1H\\
&&{}^3\mbox{He}+{}^4\mbox{He}\rightarrow{}^7\mbox{Be}+\gamma\\
&&{}^7\mbox{Be}+e^-\rightarrow{}^7\mbox{Li}+\nu\\
&&{}^7\mbox{Li}+{}^1\mbox{H}\rightarrow2\ {}^4\mbox{He}\\
&&{}^7\mbox{Be}+{}^1\mbox{H}\rightarrow2\ {}^4\mbox{He}+e^++\nu+\gamma
\end{eqnarray}

CNO cycles:
\begin{eqnarray}
&&{}^{12}\mbox{C}+{}^1\mbox{H}\rightarrow{}^{13}\mbox{C}+e^++\nu+\gamma\\
&&{}^{13}\mbox{C}+{}^1\mbox{H}\rightarrow{}^{14}\mbox{N}+\gamma\\
&&{}^{14}\mbox{N}+{}^1\mbox{H}\rightarrow{}^{15}\mbox{N}+e^++\nu+\gamma\\
&&{}^{15}\mbox{N}+{}^1\mbox{H}\rightarrow{}^{12}\mbox{C}+{}^4\mbox{He}\\
&&{}^{15}\mbox{N}+{}^1\mbox{H}\rightarrow{}^{16}\mbox{O}+\gamma\\
&&{}^{16}\mbox{O}+{}^1\mbox{H}\rightarrow{}^{17}\mbox{O}+e^++\nu+\gamma\\
&&{}^{17}\mbox{O}+{}^1\mbox{H}\rightarrow{}^{14}\mbox{N}+{}^4\mbox{He}\\
&&{}^{18}\mbox{O}+{}^1\mbox{H}\rightarrow{}^{15}\mbox{N}+{}^4\mbox{He}
\end{eqnarray}

He combustion:
\begin{eqnarray}
&&3\ {}^4\mbox{He}\rightarrow{}^{12}\mbox{C}+\gamma\\
&&{}^{12}\mbox{C}+{}^4\mbox{He}\rightarrow{}^{16}\mbox{O}+\gamma\\
&&{}^{14}\mbox{N}+{}^4\mbox{He}\rightarrow{}^{18}\mbox{O}+e^++\nu+\gamma\\
&&{}^{16}\mbox{O}+{}^4\mbox{He}\rightarrow{}^{20}\mbox{Ne}+\gamma
\end{eqnarray}

We follow thoroughly the combustion of $^2$H and $^7$Li.  Only unstable species
($^7$Be, $^{13}$N, $^{15}$O, $^{17}$F and $^{18}$F) are supposed to be at
equilibrium.  Though the main reactions of the helium burning phase have already
been implemented, we have yet to improve our code (semi-convection, equation of
state, opacity) to be able to accurately follow this phase of the evolution.

The reaction rates are computed using the analytical expressions given by
\citet{C1988}.  For the reaction $^{14}$N(p,$\gamma$)$^{15}$O, we use the
cross-section given by \citet{F2004}. A variant of the program using the NACRE
reaction rates in their approximate analytical form \citep{A1999} has been
written to facilitate the comparisons with CESAM
\citep{MoLe2005,Mo2005,MoAl2005}.

The screening factors are computed according to \citet{S1954}.

\subsection{Atmosphere}

At the last point of a stellar model Cl\'es realizes a smooth junction with a
pre-computed model atmosphere. To that end, tables giving the outer boundary
conditions in terms of $g$, $T_{e\!f\!f}$, $X$ and $Z$ have been computed from Kurucz
atmospheres \citep{K1998} at different optical depths (photo\-sphe\-re and
$\tau=1$, 10 and 100). We have also prepared tables of boundary conditions from
radiative gray Eddington atmospheres.

\subsection{Convection}

A region of a star is dynamically stable against convective movements, if
Ledoux's criterion is satisfied:
\begin{equation}
\frac{d\ln\rho}{dr}-\frac{1}{\Gamma_1}\frac{d\ln P}{dr}<0\,.
\label{eq:Ldx}
\end{equation}
In a chemically homogeneous zone, it is equivalent to the criterion of
Schwarzschild:
\begin{equation}
\nabla_{rad}<\nabla_{ad}\,,
\end{equation}
where $\nabla_{ad}=(\Gamma_3-1)/\Gamma_1$ is the adiabatic gradient.
In presence of a chemical inhomogeneity, both criteria cease to be equivalent.
In term of $\nabla$, Ledoux criterion reads
\begin{equation}
\nabla<\nabla_{Ldx}\,,
\end{equation}
where $\nabla_{Ldx}$ derives from the rewriting of equation (\ref{eq:Ldx}).
It is sufficient for our purpose to give its expression for a completely ionized
perfect gas
\begin{equation}
\nabla_{Ldx}=\nabla_{ad}+\nabla_\mu\,,
\end{equation}
where $\nabla_\mu=d\ln\mu/d\ln P$ is the gradient of the mean atomic weight.
For a non homogeneous region to be in radiative stability, 
it is not enough to satisfy Ledoux's criterion. In that case, Schwarzschild's
criterion is a necessary condition for vibrational stability. In a star, due to
nuclear processing, the
mean atomic weight generally decreases from the centre to the surface and the
term $\nabla_\mu$ is positive. We have then
\begin{equation}
\nabla_{ad}<\nabla_{Ldx}\,,
\end{equation}
and when Schwarzschild's criterion is satisfied, Ledoux's one is satisfied
too.  However, when diffusion is taken into account, gradients may develop in
small regions of the star, where this inequality is reversed and
Ledoux'criterion must be explicitly taken into account.

We have implemented the usual mixing-length theory of \citet{B1958}, also
exposed in the textbooks of \citet{C1968} and \citet{K1990}. It is a local
theory, in which the temperature gradient is obtained very simply by solving the
cubic equation
\begin{equation}
\frac{9}{4}\Gamma^3+\Gamma^2+\Gamma=A(\nabla_{rad}-\nabla_{ad})\,,
\end{equation}
where $A$ is expressed in terms of the mixing-length $\ell$, generally defined
as a multiple of the scale-height, $\ell=\alpha H_P$, with coefficient $\alpha$
chosen by the user.
\begin{equation}
A=\frac{P_T\rho}{2P_\rho P}\left[\frac{\kappa C_p\rho^2g\ell^2}
{12acT^3}\right]^2\,.
\end{equation}
The real temperature gradient is then expressed as
\begin{equation}
\nabla=\frac{\frac{9}{4}\Gamma^2\nabla_{ad}+(\Gamma+1)\nabla_{rad}}
{\frac{9}{4}\Gamma^2+\Gamma+1}\,.
\end{equation}

The variant due to \citet{H1965} has also been used for the description
of convection at low optical depth.

The Full Spectrum Turbulence (FST) theory of convection \citep{CM1991,CM1992},
has also been implemented \citep{MiMo2005}.

No effort has been made to put a grid point at the boundary of a convective
zone. Though the mixing of the material inside convective zones takes into
account the precise location of their boundaries, this results in a small
diffusion of numerical origin at the boundaries.

\paragraph{Overshooting.}

Overshooting (or undershooting) displaces the boundary of a mixed region from
$r_c$ to $r_{ov}$ given by
\begin{equation}
r_{ov}=r_c\pm\alpha_{ov}\min(H_P,h)\,,
\end{equation}
where $h$ is the thickness of the convective zone and the overshooting
coefficient $\alpha_{ov}$ is chosen by the user.

In the overshooting region, the temperature gradient can be chosen adiabatic
(only for the core),
radiative or computed according to the prescription of the mixing length theory
\citep{Go2007}.

\paragraph{Semi-convection.}

No semi-convection treatment has been implemented. For low-mass main-sequence
stars, this results in an uncertainty on the position of the boundaries of the
convective regions, at the limit of the core when it is increasing and also at
the bottom of the convective envelope when diffusion is included
\citep{Mi2005,Mo2007}.

\subsection{Diffusion}\label{diffusion}

We follow the theory of stellar diffusion developed by \citet{T1994}. From their
fundamental equations (12), (13), (18) and (19), we eliminate the electron
density and express the diffusion velocity $w_i$ in terms of the gradients of
abundances of all ions, of pressure and temperature.
\begin{equation}
w_i=\sum_ja_{ij}\frac{dX_j}{dr}+a_P\frac{d\ln P}{dr}+a_T\frac{d\ln T}{dr}\,.
\end{equation}
Though it is possible to eliminate one of the ion abundances, it seems that
keeping all the abundances ensures a better stability to the solving process.

In the standard version of the code, three groups of elements are distinguished
for the treatment of diffusion, hydrogen, helium and the metals. But in the most
advanced version of the code, the diffusion of a dozen species is considered:
the species involved in the nuclear reaction network (except Li, Be and B), Fe
and a fictitious species gathering all the other ones.  The mean degree of
ionization of each species is computed and the mean charges of the ions are used
in the computation of the diffusion coefficients.

Parametric turbulent diffusion has also been implemented \citep{MiAl2007} and
in one version of the code, radiative forces from the OP project \citep{Se2005}
are available \citep{B2006a,B2006b,B2007}.

\subsection{Mass loss}

The inclusion of a diffusion process without mass loss leads to overabundances
of hydrogen in the external layers. For solar models or stellar models with
convective envelopes, this effect is hidden by the mixture due to convection.
But for more massive stars, specially when radiative forces are taken into
account, it turns out that a slight mass loss, compatible with admitted values,
(of the order of $10^{-9}$~M$_\odot$/yr for a 10 M$_\odot$ star) is necessary
to prevent the building of unrealistic abundances in the external layers of the
star and has been implemented \citep{B2007}. Work is in progress to cope with
higher mass losses encountered in more massive stars.

%-------------------------------------------------------------------
\section{Discretization}\label{discretization}

The physical quantities describing the star are recorded at a number of discrete
points. Though their numbers can vary during the course of evolution (by
addition or deletion), these points are linked to a given material element of
the star, in other words, as long as point $P_k$ is not deleted, the value of
$m_k$ is preserved. In fact, the fundamental quantities used to characterize our
lagrangian grid are not the $m_k$ but the masses of the shells between the grid
points $m_{k+1}-m_k$, for the sake of precision in the external layers.

\subsection{Structure}

At each epoch $t$, a model is computed by solving the following difference
equations.
\begin{eqnarray}
&&P_{k+1}-P_k=\frac{1}{2}\left[
\left(\frac{dP}{dr}\right)_k+\left(\frac{dP}{dr}\right)_{k+1}\right]
(r_{k+1}-r_k)\,,\\
&&m_{k+1}-m_{k}=\frac{1}{2}(\rho_k+\rho_{k+1})(V_{k+1}-V_k)\,,\\
&&L_{k+1}-L_k=\frac{1}{2}(\epsilon_k+\epsilon_{g,k}+\epsilon_{k+1}+\epsilon_{g,k+1})
(m_{k+1}-m_k)\,,\label{eq:DeltaL}\\
&&T_{k+1}-T_k=\frac{1}{2}\left[\left(\frac{dT}{dr}\right)_k
+\left(\frac{dT}{dr}\right)_{k+1}\right](r_{k+1}-r_k)\,,
\end{eqnarray}
with $V=4\pi r^3/3$. In the energy equation (\ref{eq:DeltaL}), the $\epsilon_g$
terms involve differences between quantities at epoch $t$ and $t-\delta t$.
Omitting index $k$,
\begin{eqnarray}
\epsilon_g&=&\Bigl\{-C_v(T-T^0)+\frac{(\Gamma_3-1)C_vT}{\rho}
  (\rho-\rho^0)\nonumber\\
&&\quad-\frac{3}{2}{\mathcal R}T[2(X-X^0)+0.75(Y-Y^0)\nonumber\\
&&\quad+0.5(Z-Z^0)]\Bigr\}/\delta t\,,
\end{eqnarray}
where upper index $0$ refers to the model at epoch $t-\delta t$.

\subsection{Chemical evolution}

We update the chemical composition in a two-step process. The change of
composition due to diffusion is first computed. In a second step, the change of
composition due to nuclear burning and mixing are computed simultaneously. This
two-step process reflects the history of the development of the code (diffusion
was not included in the first versions) and will be corrected in future
versions. Fortunately, the diffusion process turns out to be fast only in the
external layers of the star where there is no nuclear burning. The code is not
able to follow the evolution of Li, Be and B.

\paragraph{Diffusion.}

In certain circumstances, the diffusion is fast\break enough in the external layers,
so that the time scale of the diffusion is smaller than any reasonable timestep
(stiff differential equations). In such cases, numerical stability of the
discrete scheme requires that the second members of the evolution equations
(\ref{eq:ChemEvol}) be calculated using the updated values.

With index $i$ referring to the chemical species, index $k$ to the grid point
and upper index $0$ to the previous epoch $t-\delta t$, our discrete equations
may be written (omitting unimportant details linked to the non equidistance of
the mesh points)
\begin{eqnarray}
&&\frac{X_{ik}-X_{ik}^0}{\delta t}=-\Bigl\{(4\pi r^2\rho X_iw_i)_{k+1/2}
-(4\pi r^2\rho X_iw_i)_{k-1/2}\Bigr\}\nonumber\\
&&\quad /(m_{k+1/2}-m_{k-1/2})\,,\\
&&w_{i,k+1/2}=\Bigl\{\sum_j a_{ij,k+1/2}(X_{j,k+1}-X_{jk})\nonumber\\
&&\quad +a_{P,k+1/2}(\ln P_{k+1}-\ln P_k)\nonumber\\
&&\quad +a_{T,k+1/2}(\ln T_{k+1}-\ln T_k)\Bigr\}/(r_{k+1}-r_k)\,,
\end{eqnarray}
where the fractional values of the spatial index have obvious meaning:
\begin{equation}
r_{k+1/2}=(r_k+r_{k+1})/2\,,\ \ldots
\end{equation}

\paragraph{Nuclear reactions.}

The life-time of the different nuclear spe\-cies are of very different orders of
magnitude, with some of them very much shorter than the timestep. We apply the
same technique as described in the previous paragraph to this stiff differential
system. As the life-time of hydrogen is of roughly the same order as the
duration of the main sequence, it is possible (and that gives better
approximation) to use an average abundance of hydrogen $\bar{X}=(X+X^0)/2$ in the
second member. This possibility has not been used in Cl\'es.
\begin{equation}
\frac{X_{ik}-X_{ik}^0}{\delta t}=\sum\limits_j R_{ijk}\,,
\label{eq:DiscNuclEvol}
\end{equation}
where the indices have the same meaning as above, except that index $0$ refers
now to the abundances obtained after the diffusion step.

In a convective zone, the $X_{ik}$ are equal for all points $k$ in the zone and
equation (\ref{eq:DiscNuclEvol}) must be replaced by the discretized form of
(\ref{eq:ChemEvolConv}). However a simple and direct discretization
would produce unphysical discontinuities when a convective zone recedes by more
than one spatial interval in one timestep. A dedicated algorithm has been
implemented to produce a continuous chemical profile in this case.

%-------------------------------------------------------------------
\section{Implementation}\label{implementation}

\subsection{The general structure}

\citet{St2006} distinguishes three different strategies for a stellar evolution
code to advance by one timestep. Our code falls in his \textit{partially
simultaneous} category.  A timestep begins with an updating of the chemical
composition followed by solving the structure equations. These two processes are
then repeated until convergence.

\paragraph{Updating the chemical composition.}

The discretized equations of evolution are nonlinear. They are solved with the
Newton-Raphson algorithm. We use the composition of the previous model
as a first guess.

\paragraph{Solving the structure equations.}

The structure equations are solved by the Newton-Raphson algorithm. When it
gives signs of difficult convergence, we apply only a fraction of the correction
suggested by the algorithm and when this recipe fails, we adopt a smaller
timestep.  In the present state of the code, we use the last computed model as a
first guess to start the iteration. We could probably improve the rate of
convergence by using an extrapolation of the last two computed models as first
guess.

\subsection{Interpolation in tables}

To guarantee the continuity of interpolated functions as well as their first
derivatives, we use cubic splines. In 1-D, continuity of the second derivative
can be achieved with cubic splines. However, this results in non-local and
sometimes unphysical behaviour of the interpolating function.  We prefer to be
able to impose the values of the first derivative at the grid points, values
that we compute locally (involving only three points), and therefore we
sacrifice the continuity of the second derivative.  This interpolation strategy
can be extended to more than one dimension in several ways.  When interpolating
physical data, affected by their inherent uncertainties, it seems that the
choice between the different variants is unimportant. The interpolation is
faster if the derivatives are stored with the values, but the tables occupy more
space in memory. Both strategies have been used in Cl\'es, we tend now to favour
the gain in computing time.

\subsection{Choice of the grid}

The grid of points of the model is adapted during the evolution in such a way as
to limit the variations of different physical quantities from one point to its
neighbours. The controlled variables are $r$, $m$, $P$ and $T$. The grid of one
model is obtained from the grid of the previous one, with addition or deletion
of points so that the following rules are satisfied:
$\Delta r\le5.10^{-3}R$, $\Delta m\le5.10^{-3}M$, $\Delta P\le5.10^{-2}P$ and
$\Delta T\le10^{-2}$.  A simple command allows the user to impose stricter or
looser rules for the grid choice.

With the default rules, a 2~M$_\odot$ model typically starts on the Hayashi
track with 700 points, reaches the zero-age main sequence with 1150 points and
keeps about the same number of points along the main sequence with local
additions and deletions.

\subsection{Choice of the timestep}

The timestep is chosen small enough to ensure that selected physical quantities
do not vary too much from one model to the next one. The controlled variables
are the local values of $T$, $P$ and $L$ and the central hydrogen abundance
$X_c$. The allowed maximum variations depend on the evolution stage and are read
from a table. This strategy gives the flexibility to adopt less severe criteria
for the pre-main sequence computation for instance. However, when the
Newton-Raphson algorithm fails to converge, the timestep is halved.  The user
has the possibility to influence, in a certain measure, the choice of the
timestep.

With no intervention of the user, a 2~M$_\odot$ model evolution needs 125 steps
for the pre-main sequence, 75 steps for the main sequence and 20 steps more for
the second gravitational contraction.

However, for evolution sequences computed with diffusion, the timestep is
effectively controlled by the necessity of convergence of the diffusion
algorithm.

\subsection{First model}

When the luminosity of the star is mainly powered by the nuclear reactions, it
is enough to give the chemical composition as a function of the mass, $X_i(m)$
to characterize a model. The good model will not be far from the model obtained
by solving the structure equation with $\epsilon_g=0$. This is not true for our
first model on the Hayashi sequence. To characterize it completely, we have to
impose, for instance, its luminosity and assume that it is contracting in a
homologous way.

%-------------------------------------------------------------------
\section{The use of Cl\'es}\label{use}

There are different levels of sophistication in the use of Cl\'es.

When it starts, Cl\'es reads a command file, i.e. a succession of commands
composed of a keyword followed by parameters, instructing the program of the
task it has to perform: mass of the star, chemical composition, EOS table,
opacity table, convection parameter,\dots\ Cl\'es supplies default values for
the parameters not specified by the user. The basic user must just know how to
write a command file to express his requirements. This is well explained in a
\textit{User's Guide} of ten or so pages.

The next step in the use of Cl\'es consists in providing the program with new
tables, such as opacities for a new metal mixture. Tools are available to help
the user in this task.

The more informed user may need to test conditions during the evolution and
interact with the program.  A subroutine called \texttt{clesuser}, which does
nothing in the standard version of the program, can be customized to meet the
need of the user. It is called by the program before each timestep and gives the
user the opportunity to read or modify the last computed model or the parameters
used for the computation.

At last, a few users have implemented in their copy of the code the features
needed for their particular research. This is facilitated by the modular
structure of the code.

%-------------------------------------------------------------------
\section{Calibration of solar models}\label{sun}
The solar models are calibrated by adjusting the mixing-length parameter
$\alpha$, the initial mass fraction $\rm X_0$ of hydrogen and the initial mass
fraction $\rm Z_0$ of heavy elements in order to reproduce, at the solar age
(4.57 $10^9$ yr, \citealt{B1995}): the observed luminosity ($3.842\;10^{33}$
$\rm erg\; s^{-1}$, \citealt{B2001}, radius ($6.9599\;10^{10}$ cm,
\citealt{A1973}) and ratio of heavy elements to hydrogen in the photosphere
(Z/X=0.0245, \citealt{G1993}). In the solar calibration we adopt a spatial and
temporal discretization finer than the standard one (see Sec.
\ref{implementation}): the number of mesh-points in a solar model is typically
$\sim2500$ and $\sim350$ time\-steps are needed to reach the solar age (including
the pre-main sequence phase).

As the values of the calibrated parameters depend on the choice of the physical
inputs, as an example we here present two solar models computed with different
opacity tables: OPAL \citep{I1996} and the recent OP opacities \citep{Se2005}
both complemented at low temperature with \citet{F2005} opacities. The remaining
common physical inputs are: the mixing-length theory for convection, GN93
\citep{G1993} metal mixture, NACRE+\citet{F2004} nuclear reaction rates and the
OPAL 2001 equation of state \citep{R2002}.  The outer boundary condition is given
by a radiative gray Eddington atmosphere computed using \citet{F2005} opacities.
The models include microscopic diffusion of H, He and Z as in the standard
version of the code (see Sec. \ref{diffusion}).

\begin{figure}
\centering
\includegraphics[width=0.4\textwidth]{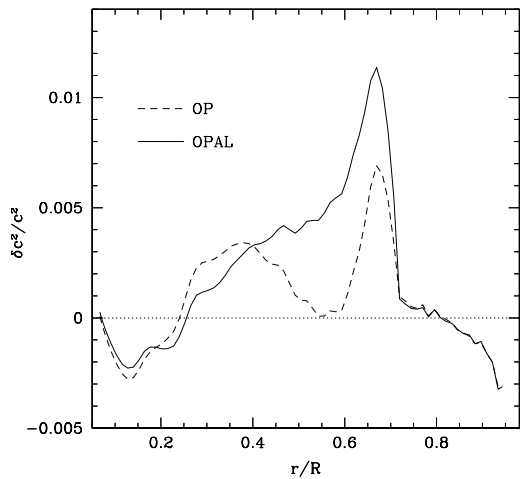}
\caption{Relative differences in the squared sound speed between the calibrated
models and the Sun \citep{Basu00}.}
\label{fig:c2}
\end{figure}

The parameters resulting from the calibration are reported in Table
\ref{tab:sun}, along with the normalised radius of the convective envelope and
the helium abundance in the envelope. The relative differences in the squared
sound speed inferred from helioseismic inversion \citep{Basu00} is shown in Fig.
\ref{fig:c2}.
 
\begin{table}[t]
\caption{Parameters of calibrated solar models. The normalized radius at the
base of the convective envelope ${\rm r_{cz}}$ and the surface helium abundance
(${\rm Y_S}$) are also reported. The values of these parameters obtained by
helioseismology are respectively $0.713 \pm 0.001$ \citep{Basu97} and
$0.245 \pm 0.005$ \citep{B2003}.}
\centering
\label{tab:sun}       % Give a unique label
\begin{tabular}{lccccc}
\hline\noalign{\smallskip}
Model & $\alpha$ & $\rm X_0$ & $\rm Z_0$ & $\rm r_{cz}$ & $\rm Y_s$ \\[3pt]
\tableheadseprule\noalign{\smallskip}
OPAL &  1.805 & 0.7043 & 0.0201 & 0.7156  &  0.2450 \\
OP   &  1.818 & 0.7056 & 0.0201 & 0.7133  &  0.2442 \\
\noalign{\smallskip}\hline
\end{tabular}
\end{table}
%==========================

%-------------------------------------------------------------------
\section{Discussion}\label{discussion}

From the standard version 18 of Cl\'es, a few users/developers working in close
collaboration in the Asteroseismology Group of Li\`ege have implemented a number
of desirable features in the code.  At the moment, there is no single copy of
the code implementing all the features described above. Our first goal is now to
gather the most significant developments already achieved and a few more in a
unique version 19. And we hope that this cycle of development will be repeated
to create new versions.

This  strategy of development was made possible because the developers have been
working in close collaboration on the same site. In less favourable
circumstances, it would require a good documentation. Unfortunately, this is not
the case, except for a user's manual.  We are still far from the ideal situation
described by \citet{H2006} but we will make the writing of a good documentation
our second priority.

%-------------------------------------------------------------------
\begin{acknowledgements}
We acknowledge financial support from the Belgian Science Policy Office (BELSPO)
in the frame of the ESA PRO\-DEX~8 program (contract C90199), from the Belgian
Interuniversity Attraction Pole (grant P5/36) and from the Fonds National de la
Recher\-che Scientifique (FNRS).
\end{acknowledgements}

%-------------------------------------------------------------------
% BibTeX users please use
%\bibliographystyle{spmpsci}

%*******************************************************************
\end{document}